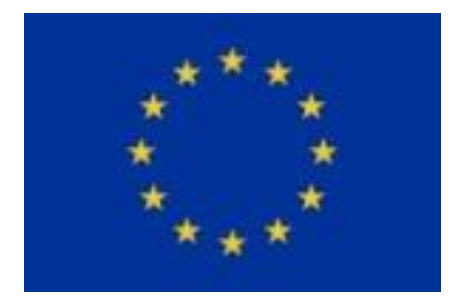

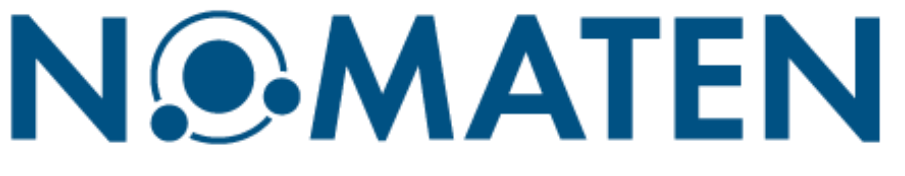

This work was carried out in whole or in part within the framework of the NOMATEN Centre of Excellence, supported from the European Union Horizon 2020 research and innovation program (Grant Agreement No. 857470) and from the European Regional Development Fund via the Foundation for Polish Science International Research Agenda PLUS program (Grant No. MAB PLUS/2018/8), and the Ministry of Science and Higher Education's initiative "Support for the Activities of Centers of Excellence Established in Poland under the Horizon 2020 Program" (agreement no. MEiN/2023/DIR/3795).

The version of record of this article, first published in Materials Today Communications, Volume 46, June 2025, 112464 (7 April 2025), is available online at Publisher's website: https://doi.org/10.1016/j.mtcomm.2025.112464

# Data-driven body-centered cubic phase prediction in cobalt-free high-entropy alloys

Xuliang Luo[a], Yulin Li[b], Tero Mäkinen[a], Silvia Bonfanti[b], Wenyi Huo[b,*], Mikko J. Alava[a,b,*]

a- Department of Applied Physics, Aalto University, P.O. Box 11000, Aalto FI-00076, Espoo, Finland

b- NOMATEN Centre of Excellence, National Centre for Nuclear Research, Otwock 05-400, Poland

***Corresponding authors:**

Mikko J. Alava: E-mail: mikko.alava@aalto.fi

Wenyi Huo: E-mail: wenyi.huo@ncbj.gov.pl

**Abstract:** High-entropy alloys (HEAs) are known for superb combination of performance attributes, making them ideal for advanced applications, e.g., nuclear engineering. The concept of cobalt-free HEAs aims to mitigate concerns about cobalt's radioactivity, however, predicting their phase formation remains challenging due to their complex compositions. In this work, we integrate six semiempirical parameters, i.e., mixing entropy ($\Delta S_{mix}$), mixing enthalpy ($\Delta H_{mix}$), atomic size difference ($\delta$), valence electron concentration ($VEC$), d-orbital energy level ($\overline{Md}$), and the $\Omega$ parameter, along with machine learning (ML) to predict the body-centered cubic phase stability in Co-free HEAs. To address the limitations of experimental data, generative adversarial networks were used to augment the dataset, thus improving the accuracy of the Gaussian process classification model used for phase prediction. After dimensionality reduction to five principal components, the model achieved an accuracy of 84%, with $\Delta H_{mix}$ and $\delta$ identified as the key descriptors influencing phase formation. This approach highlights the synergy of ML and data augmentation in accelerating the design of HEAs for advanced applications.

## 1 Introduction

Conventional alloys comprising only one or two principal elements exhibit properties that are predominantly influenced by major elements [1]. In contrast, high-entropy alloys (HEAs), which contain four or more elements in equiatomic or near-equiatomic ratios, exhibit distinctive properties due to their complex composition [2-4]. Due to their unique compositional and structural characteristics, HEAs exhibit excellent properties, such as high mechanical properties [5,6], high-temperature stability [7,8], corrosion resistance [9] and even catalytic properties [10,11], distinct from those of conventional alloys. Furthermore, the elevated radiation resistance of HEAs has garnered considerable attention, which makes them among the most promising structural materials for advanced nuclear energy system [12,13]. Vacancy-interstitial, i.e., Frenkel pairs are generated in the lattice of alloys under high-dose irradiation. Once the influence of Frenkel pairs in the lattice reaches a critical threshold, amorphization occurs [14]. HEAs can eliminate irradiation defects through a rapid crystallization process in amorphous regions under irradiation because of the complex distorted lattice of HEAs [15,16]. HEAs display the properties of delayed void formation and slow defect cluster migration, which contribute to a reduction in the size of irradiation-induced defects [17,18]. However, the majority of classical HEAs containing cobalt are not suitable for use in nuclear reactors. The elevated radioactivity and 5.27-year half-life of cobalt in high-radiation environments pose safety risks to equipment and personnel [19,20]. The HEAs containing Nb, Mo, Ti, W, Zr, V, Hf, Ta and Al fulfill the need for Co-free materials with a body-centered cubic (BCC) matrix. Moreover, refractory elements enable HEAs to withstand the extreme conditions, e.g., high temperatures [15, 21-23].

The extensive compositional space of HEAs offers prospects for property tuning. However, the intricate composition and phase compositions present a significant challenge to the conventional trial-and-error methodology [1, 24]. To reduce the cost of developing new HEA systems, the proposal of semiempirical phase formation rules has played an important role in the last decade. Zhang et al. [25] determined that solid solution phases in HEAs are prone to form when semiempirical parameters, including the entropy of mixing ($\Delta S_{mix}$), enthalpy of mixing ($\Delta H_{mix}$) and atomic size difference ($\delta$), are within the ranges from 12 to17.5 $J \cdot K^{-1} \cdot mol^{-1}$, from -15 to 5 $kJ \cdot mol^{-1}$, and less than 6.5%, respectively. The magnitude of $\Delta S_{mix}$ indicates the extent of confusion in the alloy system [26]. $\Delta H_{mix}$ reflects the miscibility of two principal elements in binary liquid alloys [27]. In accordance with the Hume–Rothery rules, the formation of solid solution phases necessitates the presence of small atomic size differences between the constituent elements [28]. The equilibrium relationships between the entropy of mixing and the enthalpy of mixing were subsequently considered, and the $\Omega$ parameter was proposed [26]. The criterion for the formation of solid solution phases in HEAs is $\Omega \geq 1.1$, and $\delta$ is corrected to be less than 6.6% on the basis of the larger data library. Through the calculation of the valance electron concentration (*VEC*) of HEAs, Guo et al. [29] delineated the stability of the FCC and BCC solid solution phases to form when $VEC \leq 6.87$ and $VEC \geq 8$, respectively. In contrast to the concept of valence electrons per atom (e/a), *VEC* takes into account the number of valence electrons, including those occupying the d-band [30,31]. Additionally, another d-electron-related parameter ($\overline{Md}$, the average value of the energy level of the d orbital), which indicates the electronegativity and atomic radius of both solute and solvent elements, was also used to predict the phase stability of HEAs [32]. The semiempirical parameters have been proven to be efficient for phase prediction in HEAs. In our previous study, a comparative analysis of the effects of these six parameters on phase stability prediction in Co-free HEAs was conducted [33,34].

The development of HEAs is a complex and time-consuming process, which is often limited by the difficulty in obtaining experimental data and the incompleteness of sample collection. In recent years, machine learning (ML) methods, as a powerful tool in data science, have shown great potential in processing and analyzing multicomponent alloy systems [1,35,36]. Combining ML with semi-empirical parameters can not only improve the prediction accuracy of phase stability in HEA systems, but also make up for the problem of insufficient experimental data [37]. However, since HEA research is still in a rapid development stage, the available data sets are usually limited, which limits the generalization ability of ML models. To address this challenge, data augmentation techniques have gained increasing attention in research involving small-scale datasets [38-39], offering substantial benefits for improving ML-based prediction tasks. Among these, generative adversarial networks (GANs), have emerged as a transformative approach for generating synthetic yet realistic data [38]. GANs consist of a generator and a discriminator that work in tandem: the generator produces synthetic samples, while the discriminator evaluates their authenticity against real data, iteratively improving the quality of the generated output [39]. They leverage a game-theoretic framework to optimize performance, generating synthetic data that enhances datasets for predictive tasks.

In this study, ML methods combined with six semi-empirical parameters, $\Delta S_{mix}$, $\Delta H_{mix}$, $\delta$, $\Omega$, $VEC$ and $\overline{Md}$, were applied to predict the stability of BCC phase in Co-free HEAs system. By constructing an ML model and applying data enhancement techniques, the formation rules of the BCC phase were more accurately identified, providing new theoretical guidance for the design of Co-free HEAs. The results of this study will help expand the application of ML in the prediction of phase stability of HEAs and provide guidance for the search of composition of high-performance Co-free HEAs.

## 2 Methods

### 2.1 Data collection

The dataset of Co-free HEAs was compiled from previously reported studies (details provided in the supplementary information). The Co-free HEAs were primarily categorized into two groups: BCC HEAs and other HEAs. Owing to the constraints of limited application conditions, the dataset comprises 226 alloy compositions.

Based on our prior research and domain knowledge of relevant features for phase prediction, we calculated the mixing enthalpy ($\Delta H_{mix}$), mixing entropy ($\Delta S_{mix}$), atomic size difference ($\delta$), and the $\Omega$ parameter, which is a parameter that takes into account the combined effects of $\Delta S_{mix}$ and $\Delta H_{mix}$, the d-orbital energy level ($\overline{Md}$), and the valence electron concentration ($VEC$). These descriptors were utilized to train the ML model, deriving them from the following equations [33,34],

$$\Delta S_{mix} = -R\sum C_i \ln C_i \quad (1),$$

$$\Delta H_{mix} = \sum_{i \neq j} C_i C_j H_{ij} \quad (2),$$

$$\delta = \sqrt{\Sigma C_i (r_i - \bar{r})^2} \quad (3),$$

$$\Omega = \frac{T_m \Delta S_{mix}}{|\Delta H_{mix}|} \quad (4),$$

$$\overline{Md} = \sum C_i (Md)_i \quad (5),$$

$$VEC = \sum C_i (VEC)_i \quad (6),$$

where $R$ = 8.314 $J \cdot K^{-1} \cdot mol^{-1}$ is the gas constant, $C_i$ is the composition fraction of element $i$, $H_{ij}$ is the mixing enthalpy between the interacting elements $i$ and $j$, $r_i$ is the atomic radius of element i, $\bar{r}$ is defined as $\bar{r} = \Sigma C_i r_i$, $T_m$ is the melting temperature of the alloy, $Md_i$ is the d-orbital energy level of element $i$, and $VEC_i$ is the valence electron concentration of element $i$.

## 2.2 Data augmentation via GANs

In this study, GANs are employed to enhance a limited dataset of HEAs to address the challenges posed by small sample sizes. Specifically, the collected dataset contains 226 HEA samples, with approximately 56% belonging to non-BCC phases. To train and evaluate the predictive models effectively, 226 samples are randomly divided into training and prediction sets, with approximately 50% in each group. The final model training set used for training is only 106 samples, and using a small dataset to train the model may result in insufficient sampling. Therefore, we first use the descriptor data of the training set to train the GAN [40,41]. The workflow of the GAN is illustrated in Figure 1. It comprises a generator that learns to approximate the data distribution of the training set, producing synthetic samples similar to real data, and a discriminator that aims to distinguish between real and synthetic samples. The adversarial interplay between these two networks enables the generator to produce high-quality fake data that closely resemble real data [42]. In this study, the GAN uses a phase structure to constrain and guide the enhancement of the target variable Co-free HEAs (simply mapping the BCC and other phase structures to 1 and 0). Each GAN-generated data point includes six fake descriptors and a structure mapping index (a continuous value between 0 and 1) indicating the predicted phase. To align with binary phase classification, the generated data are filtered: index values between -0.1 and 0.1 are classified as other phase structures (assigned 0), values between 0.9 and 1 as BCC structures (assigned 1), and values outside these ranges are discarded. The final synthetic dataset comprises 233 samples of other phase structures and 501 samples of BCC structures.

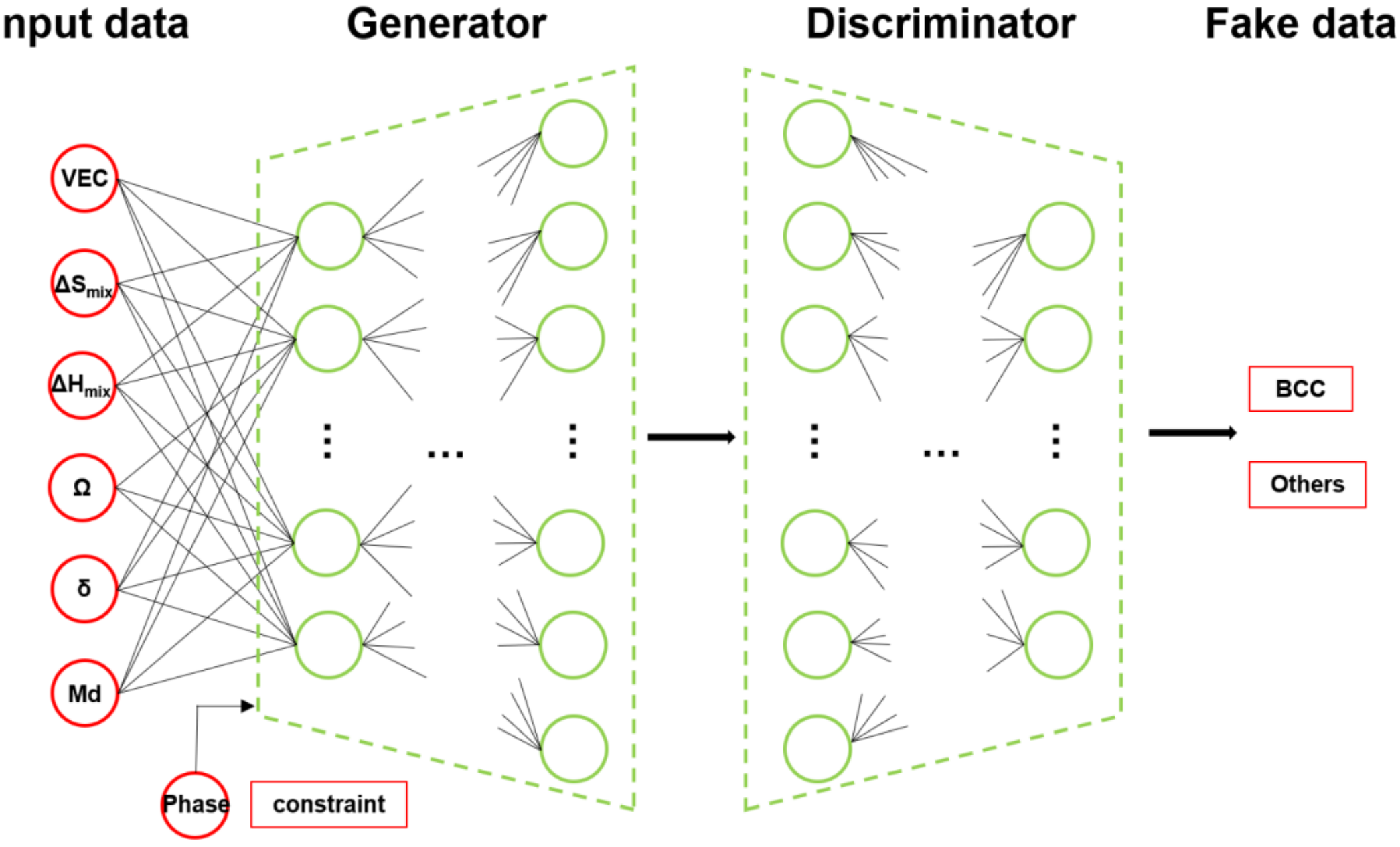

**Figure 1** Schematic diagram of the GAN model for the fake data generation of Co-free HEAs.

To verify the generated data, principal component analysis was used to perform dimensionality reduction [43] on both the experimental data, i.e., training set, and the generated data. After the two principal components were retained, a kernel smoothing function estimation plot was obtained, as shown in Figure 2a-b. Figure 2a shows the literature data, and Figure 2b shows the GAN-generated data. Density refers to the probability density estimated using kernel density estimation (KDE), which represents the smoothed distribution of data points after PCA, indicating regions of higher or lower data concentration. The generated data are distributed within the range of the literature data distribution. Furthermore, the differences between datasets are compared by calculating Wasserstein distances ($W_1$) [44] between datasets through multiple sampling calculations, and the mean Wasserstein distance between the original data and the data generated by the GAN is 0.234. This low value indicates that the data we generated are very close to the original data. Notably, The synthetic $\Delta H_{mix}$ values are generated by the GAN based on the statistical distribution of the training data, rather than calculated using physical models [33,34]. While this ensures distributional similarity (Wasserstein distance = 0.234), individual values may deviate from physically computed ones.

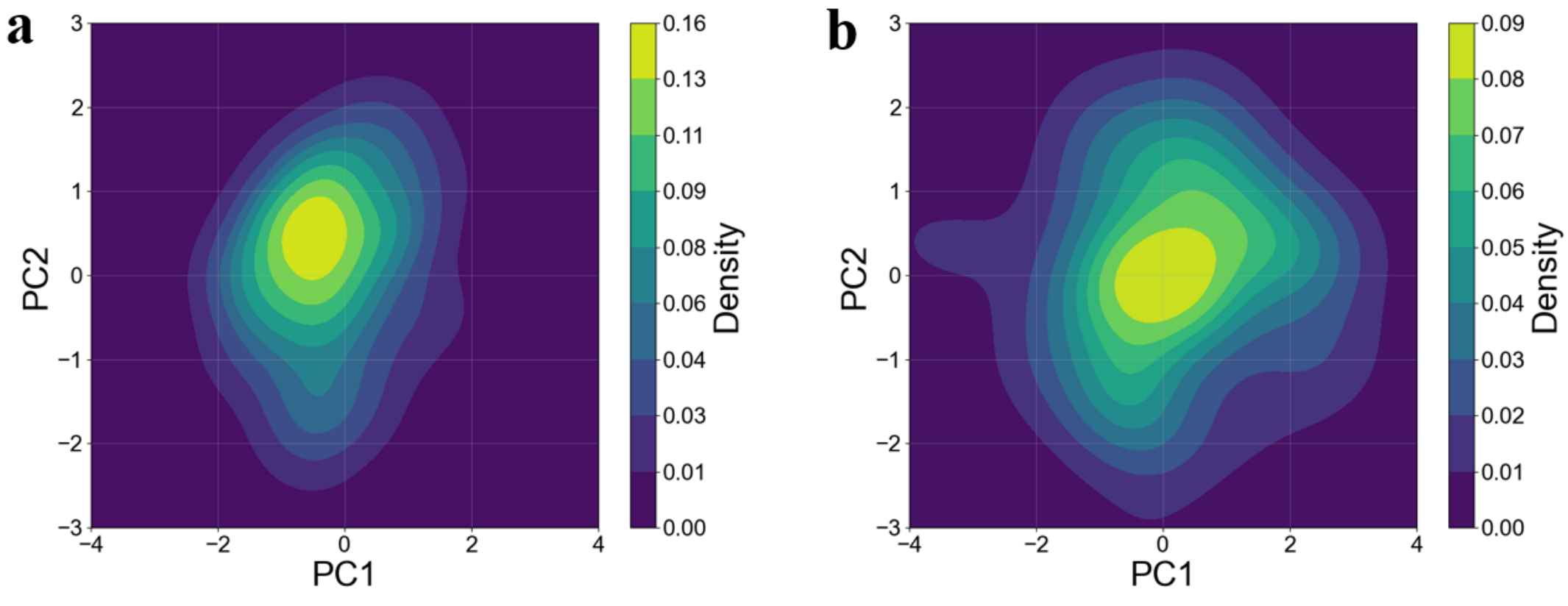

**Figure 2** 2D PCA KDE of **(a)** literature data and **(b)** generated data.

By enhancing the training data in this way, the GAN effectively improves the distribution sampling of the original dataset, enabling robust model training even with limited experimental data.

### 2.3 Bayesian optimization algorithm

The algorithm framework we use is Bayesian classification, which models the phase properties of each composition space point using surrogate functions. The surrogate function we have used is a Gaussian process (GP). Six descriptors are calculated as input feature spaces, after standardization via the formula $z_i=(x-\mu)/\sigma$ (where $z_i$ is the standardized data, $x$ is the original data point, $\mu$ is the mean of the feature, and $\sigma$ is the standard deviation of the feature). By using PCA for dimensionality reduction, it is easier to capture the main features and to combine correlated features. We use GPC to predict the posterior probability of the phase (BCC or others) for each point in the composition space. We could also do the same for other phase structures. Compared with other algorithms, GP models require fewer hyperparameters to be adjusted, so the dataset size requirement is relatively small, making them particularly suitable for resource-constrained small sample performance optimization tasks [45,46].

In the GPC algorithm, the search space is composed of selected descriptors, and after PCA dimensionality reduction, the final input variables are represented as $x = [x_1, x_2, \dots, x_n]$, where $n$ is the dimension retained after PCA dimensionality reduction. The GPC initially uses a nuisance function $f(x)$ which is a GP, at each point represented by Gaussian distribution parametrized by the mean $\mu(x)$ and standard deviation $\sigma(x)$. This nuisance function is then transformed by a logit-transform to give values between 0 and 1 which correspond to the probability estimate. The posterior mean is then computed using the Laplace approximation to yield a single valued prediction for the probability of belonging to a class, i.e. having a BCC structure with probability $P_{\mathrm{bcc}}$. The relationship between the nuisance function and the training data is calculated via a kernel function, which is a combination of an isotropic radial basis function (RBF kernel) and a constant kernel:

$$k_{RBF}(x \cdot x') = C \cdot exp\left(-\frac{\|x-x'\|^2}{2l^2}\right) \quad (7),$$

where $C$ is a constant kernel with an initial value of 1, representing the global amplitude of the kernel function; $\|x\text{-}x'\|^2$ is the Euclidean distance between $x$ and $x'$; and the parameter $l$ represents the length-scale hyperparameter, which controls the rate at which the similarity between input points $x$ and $x'$ decays as a function of their distance. The initial value of $l$ is set to 1. During optimization, $l$ is adjusted within the range from $10^{-2}$ to $10^4$. During the search process, these kernel function parameters are automatically optimized by maximizing the likelihood function to adapt to the characteristics of the data. By introducing a constant kernel, the model can adjust the strength of the correlations, whereas the RBF kernel captures the decay of local correlations in the input space. Combining this information, the GPC model achieves efficient modeling and prediction of classification problems on small sample datasets, ultimately obtaining the probability for each composition structure being BCC or other phase structure. For small sample datasets and simple binary classification problems, the GPC program is low-cost and efficient [47].

## 3. Results and discussion

### 3.1. Model training

We used a total of 840 data points, including literature data and GAN-generated data, to train the GPC model. The model utilizes six PCA-derived principal components as input features. These components are derived from the original descriptors, encapsulating the most significant variations in the dataset while minimizing redundancy. After training, it makes predictions for the BCC structure probability, $P_{\mathrm{bcc}}$, and other structure probabilities, $P_{others}$, of each original component. Figures 3a-e displays all the data kernel density estimation (KDE) plots for all the $P_{bcc}$ probabilities in the dataset with different numbers of PCA components ($n$=2-6) retained, with the X-axis representing the probabilities $P_{bcc}$ and the peak height indicating the frequency of the distribution. As shown in Figures 3a-c, the distributions are relatively broad and overlapping, indicating some uncertainty in the model's predictions. However, Figures 3d-e reveal distinct peaks, with high $P_{bcc}$ values for BCC structures and low values for other phases, indicating the model's predictions for $P_{bcc}$ are clear and well defined after 5 PCA components are retained. This is also consistent with the results in Figure 3f, where it can be seen that the first five PCA principal components retained 90% of the descriptor information.

We also trained the model without GAN-generated data. Figure S1 shows the KDE of the literature data trained with GPC. The peaks are not clearly separated, and there is considerable overlap among all the PCA components. The dataset is generally divided into a training set and a test set throughout the machine learning modeling process. After the model is trained, we use the unrelated test set only to evaluate our GPC model. As mentioned above, our test set contains a total of 121 samples, of which 81 are non-BCC phases with relatively complex structures. For all test set samples, we perform data normalization and PCA dimensionality reduction, which is consistent with the training process. The trained GPC model is then loaded to perform classification prediction on the prediction set.

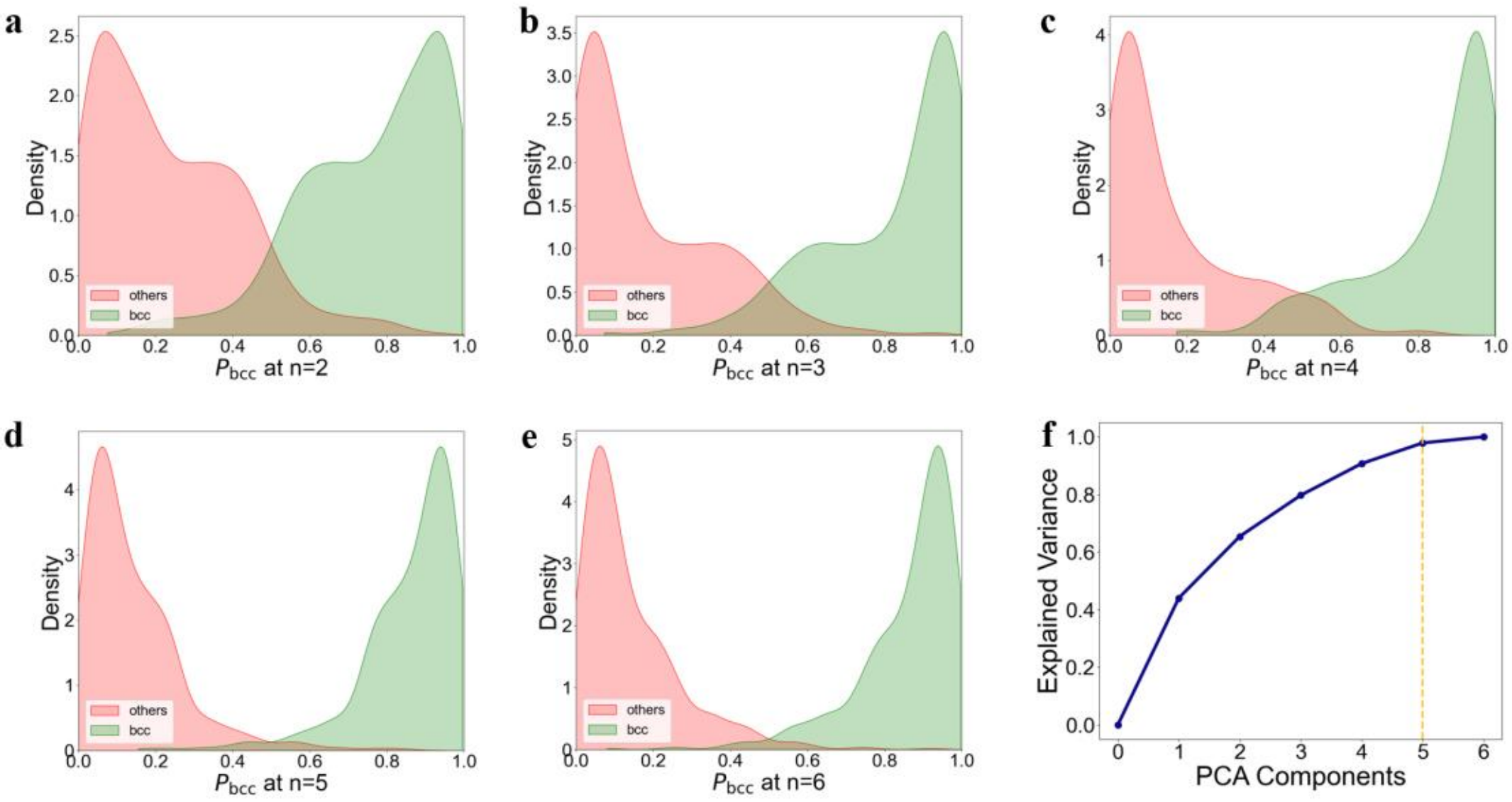

**Figure 3 (a)-(e)** Kernel density estimates of the predicted probabilities $P_{bcc}$ both for the BCC structures and others, from models trained with different PCA components retained and **(f)** PCA cumulative explained variance.

We first evaluate the model trained for each number of PCA components retained. As discussed earlier, we trained the GPC with only a literature dataset and mixed dataset (literature and generated data), and a prediction was output while retaining a different number of PCA components. Here, we calculated classification metrics on the basis of a confusion matrix to evaluate the accuracy of the model [48]. In the binary classification problem of this study, we define true positives (*TPs*) as HEAs with a BCC structure that are predicted correctly (i.e. $P_{bcc} > 0.5$), true negatives (*TNs*) as HEAs which are non-BCCs and are predicted correctly, false positives (*FPs*) as HEAs which are non-BCCs but are predicted as BCCs, and false negatives (*FNs*) as HEAs which are BCC but are predicted as other phases. Through a confusion matrix, we can calculate the true positive rate (*TPR*), true negative rate (*TNR*), false positive rate (*FPR*) and false negative rate (*FNR*) to evaluate the accuracy of sample classification prediction. These indicators are calculated via formulas (8-11):

$$TPR = \frac{TP}{TP+FN} \quad (8),$$

$$TNR = \frac{TN}{TN+FP} \quad (9),$$

$$FPR = \frac{FP}{FP+TN} \quad (10),$$

$$FNR = \frac{FN}{FN+TP} \quad (11).$$

The results are shown in Figure 4. Figure 4a-b is a prediction of the model trained with only literature data. We can observe that the decline in model accuracy with an increasing number of retained PCA components can likely be attributed to the limited size of the training dataset. In smaller datasets, retaining a higher number of PCA components often introduces noise, as these components may capture less informative or irrelevant features. This added complexity can increase the risk of overfitting, particularly when the original data distribution is sparse, as illustrated in Figure 2a. Overfitting occurs when the model focuses on noise rather than underlying patterns, reducing its generalizability. Additionally, the limited training data may lack the diversity required to adequately represent and leverage higher-dimensional features, further compromising model performance. In addition, the non-BCC phase accounts for a large proportion of the test set, and the prediction accuracy of the model trained with only literature data is extremely low for the non-BCC phase. This dilemma is significantly improved after the addition of GAN-generated data. As illustrated in Figure 4c-d, while the improvement in the TPR is relatively modest, the increased data sampling significantly enhances the model's ability to correctly identify true negatives, nearly doubling the TNR. This improvement highlights the model's enhanced ability to discern between different classes, particularly in identifying negative samples. When 5 principal components were retained in the PCA, the model achieved its highest overall accuracy of 84%. This observation aligns well with the results presented in Figure 3, further supporting the notion that an optimal number of PCA components balances the retention of informative features and the exclusion of noise, thereby improving the model predictive performance. While 84% is not at the upper limit of what might be achievable with more data or refined methods [49], it represents a strong baseline for a small-dataset scenario [50].

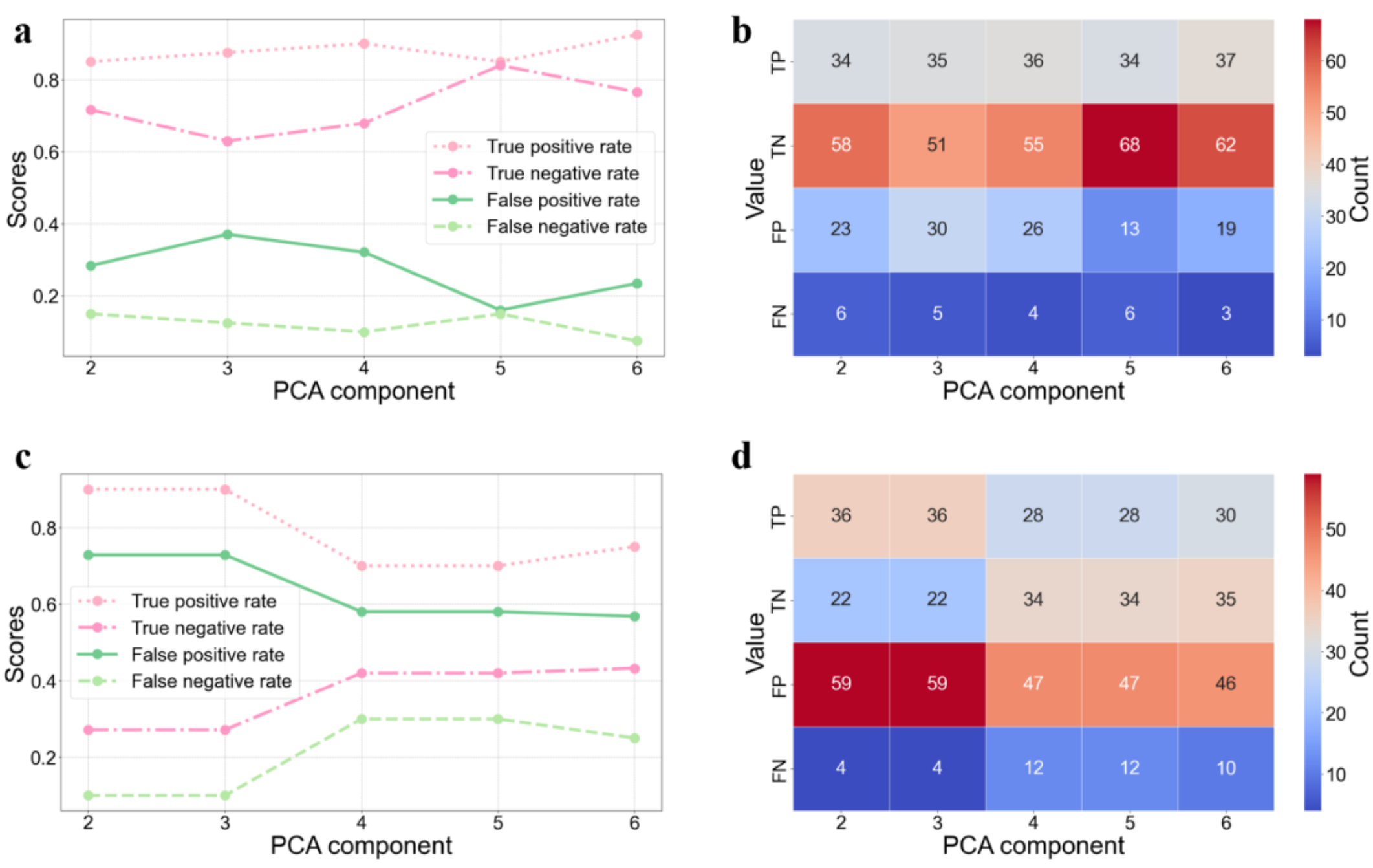

**Figure 4** Classification metric analysis and confusion matrix for the prediction results of the model trained on **(a)**, **(b)** literature data and **(c)**, **(d)** mixed data.

### 3.2 Descriptor analysis

After the above analysis, the model trained using a mixed dataset has the highest accuracy when 5 PCA principal components are retained. Therefore, we keep this PCA condition unchanged. To analyze the impact of descriptors on Co-free HEAs, the descriptors were removed one by one, and the model was retrained and predictions made. The importance of each descriptor was evaluated by calculating the performance degradation rate of the model after removing that descriptor. Figure 5 shows that removing any descriptor results in a decrease in model accuracy. Among them, $\Delta H_{mix}$ and $\delta$ have a significant effect on the accuracy of the model, whereas removing $\Delta S_{mix}$ minimizes the decrease in accuracy of the model. Such effect of an individual descriptor is consistent with our previous research [33-34,37-38,51].

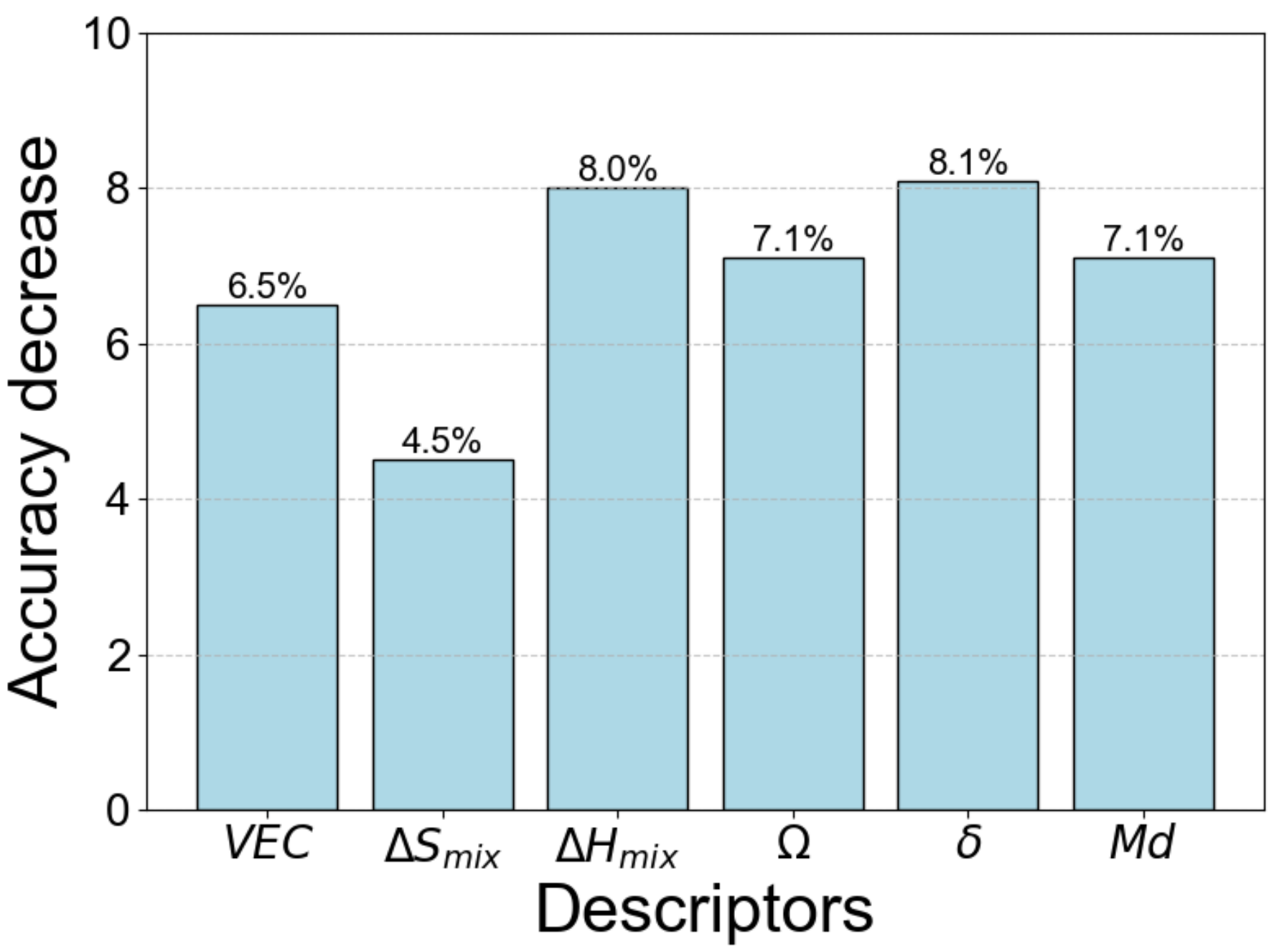

**Figure 5** Impact on the model accuracy when each descriptor is removed.

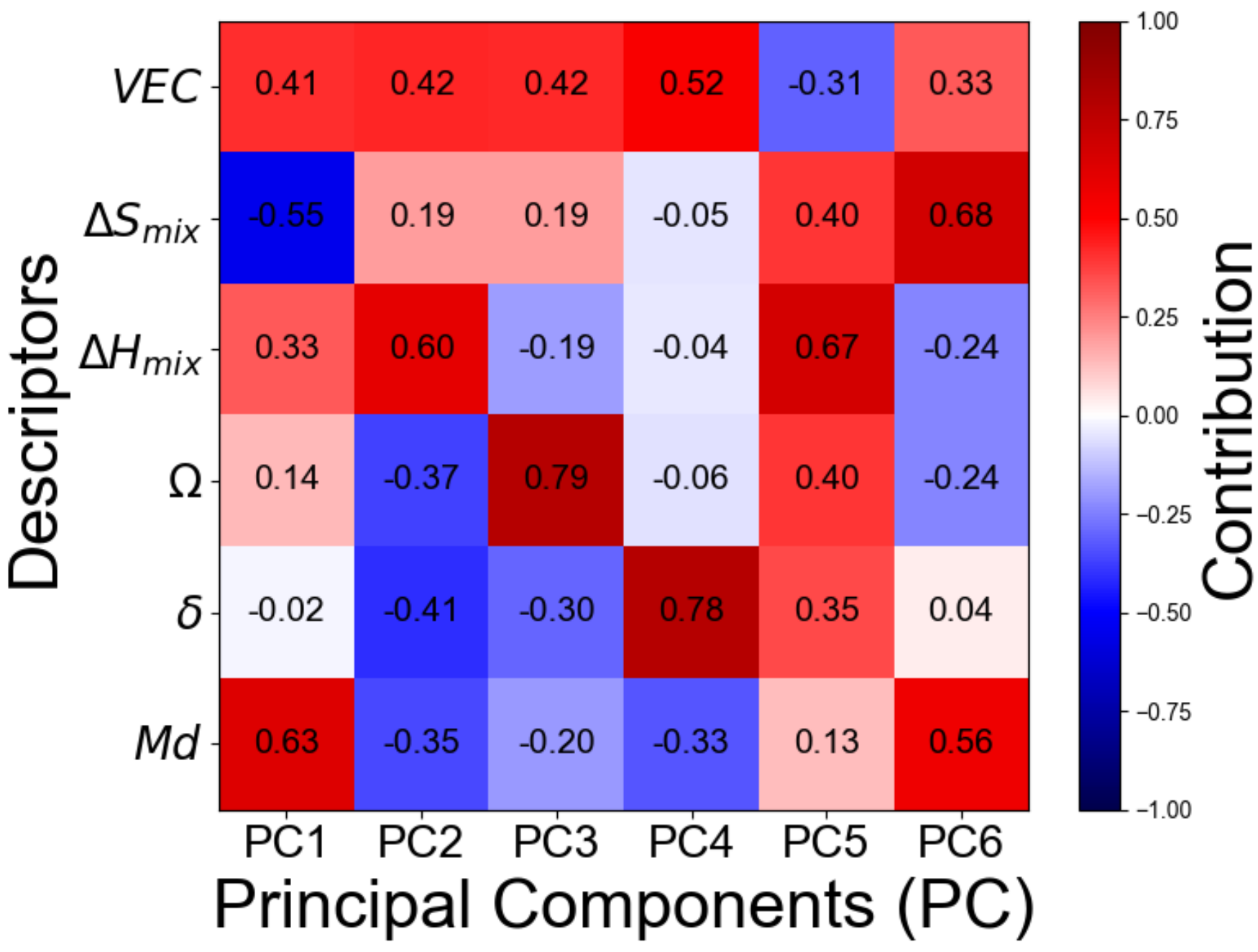

**Figure 6** PCA component weight matrix, showing the weights of the descriptors on each principal component.

Figure 6 shows the contribution of descriptors to various PCA principal components. Unlike the removal method, which reflects the overall role of descriptors in the current dataset and model, this approach includes their interaction effects with other descriptors. This method tends to focus on the independent contribution of descriptors to PCA rather than interactive or nonlinear relationships; therefore, there may be differences in the results between the two evaluation methods. This method tends to focus on the independent contribution of descriptors to PCA rather than interactive or nonlinear relationships. Therefore, there may be differences in the results between the two evaluation methods. The distinct contributions to PC1 are primarily driven by $\overline{Md}$, $VEC$ and $\Delta H_{mix}$, which positively affect PC1, whereas $\delta$ shows a negative contribution. This suggests that higher $\delta$ values align with lower PC1 coordinates. If regions of low PC1 values in the PCA space correspond to a higher probability of BCC phase formation, i.e., $P_{\mathrm{bcc}}$, then a smaller atomic size difference could indeed favor the BCC phase. According to the Hume–Rothery rules, atomic size differences significantly influence the formation and stability of solid solution phases. When the atomic size difference between solute and solvent atoms is large, it can lead to lattice distortion, introducing strain energy that destabilizes the crystal structure. Furthermore, significant lattice distortion can impede atomic diffusion during solidification, which may promote the formation of intermetallic compounds rather than solid solution phases such as the BCC structure. In extreme cases, the reduced mobility of atoms due to such lattice distortion may even favor the formation of amorphous phases [28]. These findings highlight that while parameters such as atomic radius and mixing enthalpy are not solely decisive, they significantly influence the compositional design of Co-free HEAs. Furthermore, the integration of data-augmented machine learning has demonstrated strong predictive accuracy in forecasting phase formation in these alloys.

## 4. Conclusion

In this work, we used a GAN to generate data close to the literature data and, combined with PCA dimensionality reduction, trained a GPC model that can be used to predict the phase structure of Co-free HEAs. After combining the GAN-enhanced data, the model's prediction accuracy for the BCC phase structure is significantly improved. The model has a maximum accuracy of 84% when 5 PCA principal components are retained. We also discussed the six descriptors used, and the results showed that $\Delta H_{mix}$ and $\delta$ have a greater impact on model accuracy. These findings align with previous studies and experimental data, reinforcing the importance of these descriptors in phase stability prediction. These results demonstrate the potential of combining machine learning and data augmentation techniques to accelerate the rational design of HEAs. This approach offers a robust framework for predicting phase stability in complex alloy systems. Future enhancements, e.g., integrating larger datasets, physics-informed constraints, or ensemble methods, could elevate both prediction accuracy and material discovery.

## Declaration of competing interests

The authors declare that they have no known competing financial interests or personal relationships that could have appeared to influence the work reported in this paper.

## Data availability

The data will be made available upon request.

## CRediT author statement

Xuliang Luo: Investigation, Formal analysis, Visualization, Software, Writing – original draft. Yulin Li: Investigation, Formal analysis, Writing – original draft. Tero Mäkinen: Formal analysis, Methodology, Software, Writing – review & editing. Silvia Bonfanti: Formal analysis, Software, Writing – review & editing. Wenyi Huo: Conceptualization, Methodology, Writing – original draft, Writing – review & editing, Supervision. Mikko J. Alava: Methodology, Software, Writing – review & editing, Supervision.

## Acknowledgments

X.L. acknowledges the support from the Finnish Ministry of Education and Culture's Pilot for Doctoral Programs (Pilot project Mathematics of Sensing, Imaging and Modelling). Y.L., S.B., W.H. and M.A. acknowledge the support from European Union Horizon 2020 research and innovation program under grant agreement no. 857470, and European Regional Development Fund via the Foundation for Polish Science International Research Agenda PLUS program grant no. MAB PLUS/2018/8. The publication was created within the framework of the project of the Minister of Science and Higher Education "Support for the activities of Centres of Excellence established in Poland under Horizon 2020" under contract no. MEiN/2023/DIR/3795. T.M. and M.A. acknowledge funding from FinnCERES flagship (151830423), Business Finland (211835, 211909, 211989), and Future Makers programs. M.A. acknowledges funding from the Finnish Cultural Foundation. Aalto Science-IT project is acknowledged for computational resources.

## References

[1] C. Wen, Y. Zhang, C. Wang, D. Xue, Y. Bai, S. Antonov, L. Dai, T. Lookman, Y. Su, Machine learning assisted design of high entropy alloys with desired property, Acta Mater. 170 (2019) 109-117.
[2] B. Cantor, I.T.H. Chang, P. Knight, A.J.B. Vincent, Microstructural development in equiatomic multicomponent alloys, Mater. Sci. Eng. A 375-377 (2004) 213-218.
[3] J.W. Yeh, S.K. Chen, S.J. Lin, J.Y. Gan, T.S. Chin, T.T. Shun, C.H. Tsau, S.Y. Chang, Nanostructured high-entropy alloys with multiple principal elements: novel alloy design concepts and outcomes, Adv. Eng. Mater. 6 (2004) 299-303.
[4] W. Huo, F. Fang, H. Zhou, Z. Xie, J.K. Shang, J. Jiang. Remarkable strength of CoCrFeNi high-entropy alloy wires at cryogenic and elevated temperatures. Scr. Mater. 141 (2017) 125-128.
[5] A. Olejarz, W. Huo, M. Zieliński, R. Diduszko, E. Wyszkowska, A. Kosińska, D. Kalita, I. Jóźwik, M. Chmielewski, F. Fang, Ł. Kurpaska. Microstructure and mechanical properties of mechanically-alloyed CoCrFeNi high-entropy alloys using low ball-to-powder ratio. J. Alloy. Comp. 938 (2023) 168196.
[6] Y. Chen, X. An, S. Zhang, F. Fang, W. Huo, P. Munroe, Z. Xie. Mechanical size effect of eutectic high entropy alloy: Effect of lamellar orientation, J. Mater. Sci. Technol. 82 (2021) 10-20.
[7] S. Praveen, H.S. Kim. High-entropy alloys: Potential candidates for high-temperature applications - an overview, Adv. Eng. Mater. 20 (2018) 1700645.
[8] Q. Xu, K. Karimi, A.H. Naghdi, W. Huo, C. Wei, S. Papanikolaou, Nanoindentation responses of NiCoFe medium-entropy alloys from cryogenic to elevated temperatures. J. Iron Steel Res. Int. 31 (2024) 2068-2077.

[9] W. Huo, S. Wang, F. Fang, S. Tan, Ł. Kurpaska, Z. Xie, H. S. Kim and J. Jiang, Microstructure and corrosion resistance of highly <111> oriented electrodeposited CoNiFe medium-entropy alloy films. J. Mater. Res. Technol. 20 (2022) 1677-1684.
[10] S. Wang, H. Yan, W. Huo, A. Davydok, M. Zając, J. Stępień, H. Feng, Z. Xie, J.K. Shang, P.H.C. Camargo, J. Jiang, F. Fang. Interstitial-atom-induced multiple nano-twinned high entropy alloy catalysts for efficient water electrolysis, Appl. Catal. B Environ. Energy 363 (2025) 124791.
[11] W. Huo, S. Wang, F. J. Dominguez-Gutierrez, K. Ren, Ł. Kurpaska, F. Fang, S. Papanikolaou, H.S. Kim and J. Jiang, High-entropy materials for electrocatalytic applications: a review of first-principles modeling and simulation. Mater. Res. Lett. 11 (2023) 713-732.
[12] C.M. Barr, J.E. Nathaniel, K.A. Unocic, J. Liu, Y. Zhang, Y. Wang, M.L. Taheri, Exploring radiation induced segregation mechanisms at grain boundaries in equiatomic CoCrFeNiMn high entropy alloy under heavy ion irradiation, Scr. Mater. 156 (2018) 80-84.
[13] M. Moschetti, P.A. Burr, E. Obbard, J.J. Kruzic, P. Hosemann, B. Gludovatz, Design considerations for high entropy alloys in advanced nuclear applications, J. Nucl. Mater. 567 (2022) 153814.
[14] W. Guo, T. Iwashita, T. Egami, Universal local strain in solid-state amorphization: The atomic size effect in binary alloys, Acta Mater. 68 (2014) 229-237.
[15] T. Egami, M. Ojha, O. Khorgolkhuu, D.M. Nicholson, G.M. Stocks, Local electronic effects and irradiation resistance in high-entropy alloys, JOM 67 (2015) 2345-2349.
[16] T. Nagase, P.D. Rack, J.H. Noh, T. Egami, In-situ TEM observation of structural changes in nano-crystalline CoCrCuFeNi multicomponent high-entropy alloy (HEA) under fast electron irradiation by high voltage electron microscopy (HVEM), Intermetallics 59 (2015) 32-42.
[17] E.J. Pickering, A.W. Carruthers, P.J. Barron, S.C. Middleburgh, D.E.J. Armstrong, A.S. Gandy, High-entropy alloys for advanced nuclear applications, Entropy 23 (2021) 98.
[18] S.Q. Xia, X. Yang, T.F. Yang, S. Liu, Y. Zhang, Irradiation resistance in $Al_xCoCrFeNi$ high entropy alloys, JOM 67 (2015) 2340-2344.
[19] N.A.P.K. Kumar, C. Li, K.J. Leonard, H. Bei, S.J. Zinkle, Microstructural stability and mechanical behavior of FeNiMnCr high entropy alloy under ion irradiation, Acta Mater. 113 (2016) 230-244.
[20] M. Luebbe, J. Duan, F. Zhang, J. Poplawsky, H. Pommeranke, M. Arivu, A. Hoffman, M. Buchely, H. Wen, A high-strength precipitation hardened cobalt-free high-entropy alloy, Mater. Sci. Eng. A 870 (2023) 144848.
[21] P.P. Cao, H.L. Huang, S.H. Jiang, X.J. Liu, H. Wang, Y. Wu, Z.P. Lu, Microstructural stability and aging behavior of refractory high entropy alloys at intermediate temperatures, J. Mater. Sci. Technol. 122 (2022) 243-254.
[22] J.L. Zhou, Y.H. Cheng, Y.X. Chen, X.B. Liang, Composition design and preparation process of refractory high-entropy alloys: A review, Int. J. Refract. Met. Hard Mater. 105 (2022) 105836.

[23] W. Huo, H. Shi, X. Ren, J. Zhang, Microstructure and wear behavior of CoCrFeMnNbNi high-entropy alloy coating by TIG cladding, Adv. Mater. Sci. Eng. (2015) 647351.
[24] D.B. Miracle, O.N. Senkov, A critical review of high entropy alloys and related concepts, Acta Mater. 122 (2017) 448-511.
[25] Y. Zhang, Y.J. Zhou, J.P. Lin, G.L. Chen, P.K. Liaw, Solid-solution phase formation rules for multi-component alloys, Adv. Eng. Mater. 10 (2008) 534-538.
[26] X. Yang, Y. Zhang, Prediction of high-entropy stabilized solid-solution in multi-component alloys, Mater. Chem. Phys. 132 (2012) 233-238.
[27] A.K. Niessen, F.R. De Boer, R.D. Boom, P.F. De Châtel, W.C.M. Mattens, A.R. Miedema, Model predictions for the enthalpy of formation of transition metal alloys II, Calphad 7 (1983) 51-70.
[28] T. Egami, Y. Waseda, Atomic size effect on the formability of metallic glasses, J. Non-Cryst. Solids 64 (1984) 113-134.
[29] S. Guo, C. Ng, J. Lu, C.T. Liu, Effect of valence electron concentration on stability of fcc or bcc phase in high entropy alloys, J. Appl. Phys. 109 (2011) 103505.
[30] S. Guo, Phase selection rules for cast high entropy alloys: an overview, Mater. Sci. Technol. 31 (2015) 1223-1230.
[31] M.G. Poletti, L. Battezzati, Electronic and thermodynamic criteria for the occurrence of high entropy alloys in metallic systems, Acta Mater. 75 (2014) 297-306.
[32] M. Morinaga, N. Yukawa, H. Ezaki, H. Adachi, Solid solubilities in transition-metal-based fcc alloys, Philos. Mag. A 51 (1985) 223-246.
[33] Y. Li, Ł. Kurpaska, E. Lu, Z. Xie, H.S. Kim, W. Huo, Body-centered cubic phase stability in cobalt-free refractory high-entropy alloys, Res. Phys. (2024) 107688.
[34] Y. Li, A. Olejarz, Ł. Kurpaska, E. Lu, M.J. Alava, H.S. Kim, W. Huo, Designing cobalt-free face-centered cubic high-entropy alloys: A strategy using d-orbital energy level, Int. J. Refract. Hard Mat. 124 (2024) 106834.
[35] W. Huang, P. Martin, H.L. Zhuang, Machine-learning phase prediction of high-entropy alloys, Acta Mater. 169 (2019) 225-236.
[36] Z. Rao, P.Y. Tung, R. Xie, Y. Wei, H. Zhang, A. Ferrari, T.P.C. Klaver, F. Körmann, P.T. Sukumar, A.K. da Silva, Y. Chen, Z. Li, D. Ponge, J. Neugebauer, O. Gutfleisch, S. Bauer, D. Raabe, Machine learning-enabled high-entropy alloy discovery, Science, 378 (2022) 78-85.
[37] W. Zhu, W. Huo, S. Wang, X. Wang, K. Ren, S. Tan, F. Fang, Z. Xie, J. Jiang, Phase formation prediction of high-entropy alloys: a deep learning study, J. Mater. Res. Technol. 18 (2022) 800-809.
[38] W. Zhu, W. Huo, S. Wang, Ł. Kurpaska, F. Fang, S. Papanikolaou, H.S. Kim, J. Jiang, Machine learning based hardness prediction of high-entropy alloys for laser additive manufacturing, JOM 75 (2023) 5537-5548.
[39] I. Goodfellow, J. Pouget-Abadie, M. Mirza, B. Xu, D. Warde-Farley, S. Ozair, A. Courville, Y Bengio, Generative adversarial nets, Proc. Adv. Neural. Inf. Procc. Syst. 27 (2014).
[40] Z. Zhou, Y. Shang, X. Liu, Y. Yang, A generative deep learning framework for inverse design of compositionally complex bulk metallic glasses, npj Comput. Mater.

9 (2023) 15.
[41] L. Zi, WT. Nash, SP. O'Brien, Y. Qiu, RK. Gupta, N. Birbilis, cardiGAN: A generative adversarial network model for design and discovery of multi principal element alloys. J. Mater. Sci. Technol. 125 (2022) 81-96.
[42] C. Chen, H. Zhou, W. Long, G. Wang, J. Ren, Phase prediction for high-entropy alloys using generative adversarial network and active learning based on small datasets. Sci. China Technol. Sci. 12 (2023) 3615-3627.
[43] I.Y. Miranda-Valdez, T. Mäkinen, S. Coffeng, A. Päivänsalo, L. Jannuzzi, L. Viitanen, J. Koivisto, M.J. Alava, Accelerated design of solid bio-based foams for plastics substitutes. Mater. Horiz. 12 (2025) 1855-1862.
[44] C. Villani, Optimal transport: old and new. Berlin: Springer. 388 (2009) 23.
[45] V. Torsti, T. Mäkinen, S. Bonfanti, Koivisto J, M.J. Alava, Improving the mechanical properties of Cantor-like alloys with Bayesian optimization. APL Mach. Learn. 2 (2024) 016119.
[46] B. Vela, D. Khatamsaz, C. Acemi, I. Karaman, R. Arroyave, Data-augmented modeling for yield strength of refractory high entropy alloys: A bayesian approach. Acta Mater. 261 (2023) 119351.
[47] S. Lee, S. Byeon, H. S. Kim, H. Jin, S. Lee, Deep learning-based phase prediction of high-entropy alloys: Optimization, generation, and explanation. Mater. Des. 197 (2021) 109260.
[48] A. Luque, A. Carrasco, A. Martin, A. Heras, The impact of class imbalance in classification performance metrics based on the binary confusion matrix. Pattern Recognit. 91 (2019) 216-231.
[49] Q. Han, Z. Lu, S. Zhao, Y. Su, H. Cui, Data-driven based phase constitution prediction in high entropy alloys. Comp. Mater. Sci. 215 (2022) 111774.
[50] S. Hou, M. Sun, M. Bai, D. Lin, Y. Li, W. Liu, A hybrid prediction frame for HEAs based on empirical knowledge and machine learning. Acta Mater. 228 (2022) 117742.
[51] Y. Li, H. Yan, S. Wang, X. Luo, Ł. Kurpaska, F. Fang, J. Jiang, H.S. Kim, W. Huo, Toward predictable phase structures in high-entropy oxides: A strategy for screening multicomponent compositions. Mater. Des. 248 (2024) 113497.